# Tuner control of spoke012 SRF cavity

# for C-ADS injector Ⅰ at IHEP *

LIU Na(刘娜)[1;2;1] SUN Yi(孙毅)[2] WANG Guang-Wei(王光伟)[2]

MI Zheng-Hui (米正辉)[1;2] LIN Hai-Ying(林海英)[2] WANG Qun-Yao(王群要)[2]

1 University of Chinese Academy of Sciences, Beijing 100049, China
2 Institute of High Energy Physics, CAS, Beijing 100049, China

**Abstract**：A new tuner control system of spoke superconducting radio frequency (SRF) cavity has been developed and applied to cryomodule I (CM1) of C-ADS injector I at IHEP. We have successfully implemented the tuner controller and achieved a cavity tuning phase error of ±0.7° (peak to peak) in the presence of electromechanical coupled resonance. This paper will present the preliminary experimental results based on the new tuner controller under proton beam commissioning.

**Key words**：  tuner control, spoke cavity, electromechanical coupled resonance

**PACS:** 29.20.Ej

## 1. Introduction

The China Accelerator Driven sub-critical System (C-ADS) is a proposed project to solve the nuclear waste problem and safety operation for nuclear power plant in China. The injector I is a part of 25MeV main linac for the C-ADS project facility. It will be operated in the continuous wave (CW) mode and provide 10MeV proton beam as well [1]. The first spoke SRF cavity, whose frequency is 325MHz and β=0.12 (spoke012), has been developed by Institute of High Energy Physics (IHEP) Chinese Academy of Sciences. And the first beam commissioning was carried out in September 2015.

The spoke012 SRF cavity is chosen to accelerate proton in the C-ADS injector I. The CM1 of injector I, which consists of seven spoke012 cavities, has been constructed at IHEP. In order to control the resonant frequency of each cavity to 325MHz, a new tuner control system has been developed to achieve a cavity tuning phase error of ±0.7° (peak to peak). It also provides Graphical User Interface (GUI) based on Control System Studio (CSS) interface based on Experimental Physics and Industrial Control System (EPICS) for remote operation.

The spoke012 SRF cavities have high Q in consequence the cavities have to be operated with narrow bandwidths. The electromechanical coupled resonance (ponderomotive instability) results in the amplitude and phase instability of field in superconducting cavity [2]. The resonance frequency appears in 200-250Hz within ±3dB bandwidths of spoke012 cavities. The fast feedback control of piezo electric mechanical tuners has been used with some success to tune the cavity frequency and stabilize the field in SRF cavity [3]. This paper will present results of detuning compensation for 325MHz spoke012 SRF cavity in injector I of C-ADS project at IHEP under proton beam test.

*Supported by Proton linac accelerator I of China Accelerator Driven sub-critical System (Y12C32W129)
1) E-mail: liun@ihep.ac.cn

## 2. The mechanical structure of tuner

The cavity tuners are mechanical devices designed to tune the cavity frequency. As Fig. 2 (a) and (b) show, a mechanical tuner using a stepping motor as a slow tuner and piezo actuators as a fast tuner is used to compensate for the cavity detuning.

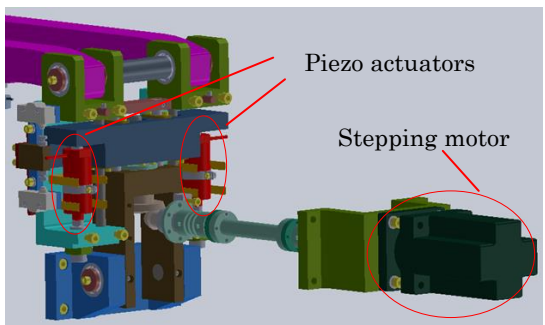

Fig. 2. (a) The 3D graph of 325MHz spoke012 cavity tuner;

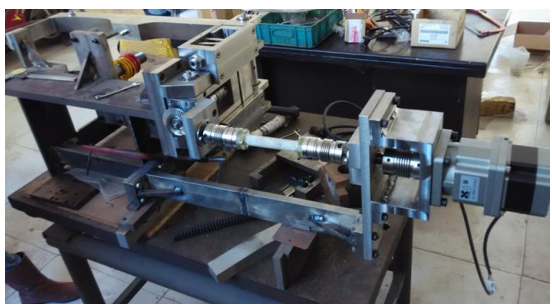

Fig. 2. (b) The 325MHz spoke012 cavity tuner on the test bed.

The mechanical stepping motor tuner works at room temperature for a large adjustable frequency range about

±200kHz and the piezo tuner works at 2K environment for a small but rapid frequency adjustment. The main parameters of 325MHz tuner are listed in Table 1[4].

Table 1. Main parameters of 325MHz tuner.

| Type | Stepping Motor | Piezo |
|---|---|---|
| Tuning rate | slow | fast |
| Operating temperature | room temperature | 2K |
| Tuning range | ±20kHz | ±2kHz |
| Cavity tuning sensitivity | 1kHz/um | 40nm/V |
| Frequency sensitivity | 0.33Hz/pulse | 0.04Hz/mV |
| Resolution | um | nm |
| Harmonic drive ratio | 1:50 | -- |

The step angle of stepping motor (PKE596AC-HS50) is 0.009°/step which has been 80 subdivided by adjusting the motor driver. The reduction ratio of reduction gears is 1:3. The screw pitch of tuner is 2 mm. The theory formula of pulse number and cavity frequency shift is given by:

$$\Delta f = pn \times \frac{0.009^0}{360^0} \times \frac{1}{50} \times \frac{1}{3} \times 2 \times 1. \quad (1)$$

$\Delta f$: cavity frequency shift; $pn$: pulse numbers.

According to Formula (1), the theoretical resolution of main stepping motor is 0.33nm/pulse. But the actual resolution is far lower than theoretical resolution due to installation errors and many other factors [4]. Fig. 3 shows the pulse number of stepping motor vs. cavity frequency curve at Eacc=1MV/m. The return difference caused by the backlash of gears has been shown inside the red circle. As the maximum backlash value of stepping motor is about 337Hz close to the full bandwidths of cavity (~400Hz).

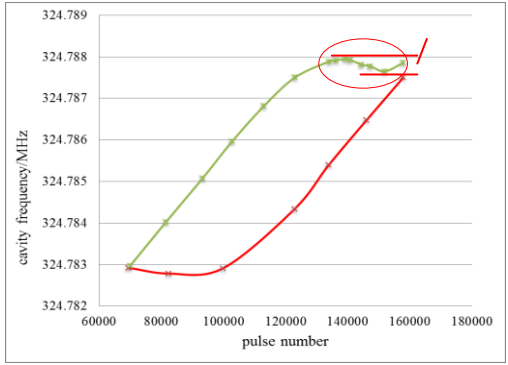

Fig. 3. Pulse number of stepping motor vs. cavity frequency.

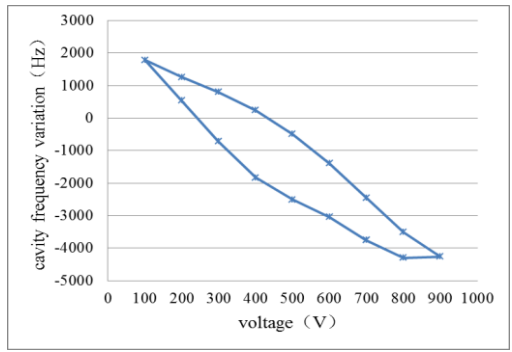

Fig. 4. (a) Excitation voltage of one piezo actuator vs. cavity frequency variation at room temperature;

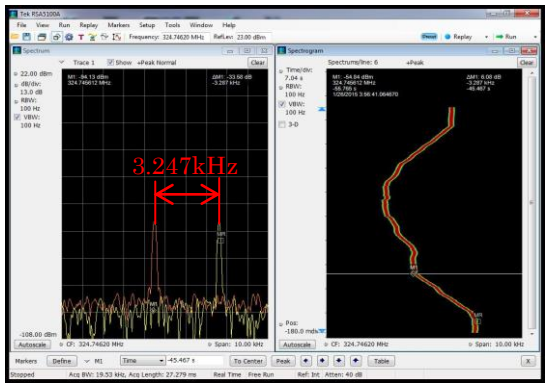

Fig. 4. (b) The cavity frequency variation of two piezo actuators at 2K temperature.

As Fig. 4 (a) and (b) show, the cavity

frequency has changed about 6kHz by one piezo actuator at room temperature during the excitation voltage varies from 0 to 900 V (the offset is 500V). However the adjustable range of one piezo actuator decreases to 1.63kHz at 2K temperature. And furthermore, the hysteresis phenomenon is very evident to motor tuner and piezo tuner during round trip. Large frequency detuning but narrow bandwidths cavity remains a big challenge for the precision of tuner control system.

## 4. Resonance stabilization using feedback control

### 4.1 The main factors influencing the cavity frequency

The cavity frequency is disturbed by many factors, such as the Lorentz force, the pressure of the liquid helium, beam loading and microphonics. The main factors of detuning vary from cavity to cavity in an operational machine. The main RF parameters of the spoke012 cavity have been listed in Table2.

Table 2. Main RF parameters of cavity.

| Parameter | Design | Measured(2K) |
|---|---|---|
| Frequency | 325MHz | 324.75MHz |
| $Q_L$ | 7.50E+05 | >7.5E+05 |
| Beam current (pulse operation) | 10mA | 10mA |
| Eacc (in operation) | 6.08MV/m | 5~10MV/m |
| Bandwidth(±3dB) | 433Hz | ~ 400Hz |
| Cavity Tuning Sensitivity | 1kHz/um | 1.1kHz/um |
| Factor of LFD | -4Hz/(MV/m)^2 | ~ -10~-18 Hz/(MV/m)^2 |
| Frequency Shift caused by LFD | -144Hz | -370~665Hz |
| Mechanical Vibration Modes | -- | 200-250 |

To the low-velocity proton

accelerators, the beam loading was often negligible. The cryomodule works at 2K environment and the fluctuation of liquid helium pressure (LHe) is very small. Any cavity will have an infinite number of mechanical eigenmodes of vibration, which can be driven by microphonics and Lorentz force, to cause the frequency shift. On the other hand, the coupling between the electromagnetic modes and the mechanical modes can lead to resonance instabilities. Therefore the issues associated with ponderomotive instability and microphonics are the major challenges for tuning the cavity and stabilizing the RF fields in superconducting cavity.

The mechanical eigenmodes of vibration on one of spoke012 cavities with turbo and scroll pump can be seen in Fig. 8. The main resonance frequency appears in ~250Hz and the transfer function is very complex due to the large number of low-frequency modes.

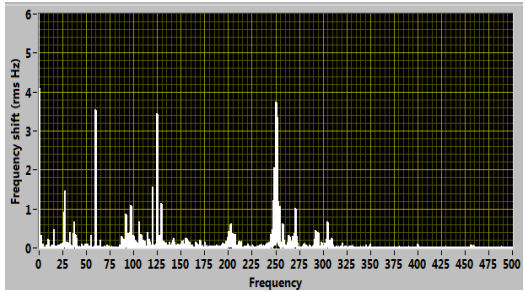

Fig.8. Mechanical eigenmodes of vibration for one of spoke012 cavities with turbo and scroll pump.

According the adiabatic theorem, the frequency shift $\Delta\omega_\mu$ caused by mechanical mode $\mu$ is given by

$$\ddot{\Delta\omega}_\mu + \frac{2}{\tau_\mu}\dot{\Delta\omega}_\mu + \Omega_\mu^2\omega_\mu = -k_\mu\Omega_\mu^2 V^2 + n(t). \quad (2)$$

$\tau_\mu$ : the decay time of mechanical mode $\mu$ ; $\Omega_\mu$ : the frequency of mechanical mode $\mu$ ; $k_\mu$ : the coupling between the RF field and mechanical mode $\mu$ ; $V$ : the field amplitude; $n(t)$ : an additional driving term representing external vibrations or microphonics. In steady-state, the total frequency shift is

$$\Delta\omega_0 = \sum_\mu \Delta\omega_{\mu 0} = -V^2 \sum_\mu k_\mu. \quad (3)$$

$\sum_\mu k_\mu$ is the static Lorentz coefficient of the cavity [7].

Eq.(2) describes how the cavity frequency is affected by the field amplitude and microphonics. Eq.(3) illustrates that the total Lorentz function for a cavity is the sum of all the functions for the individual mechanical modes. The Lorentz coefficient of 7# spoke012 cavity has been measured and shown in Fig. 10.

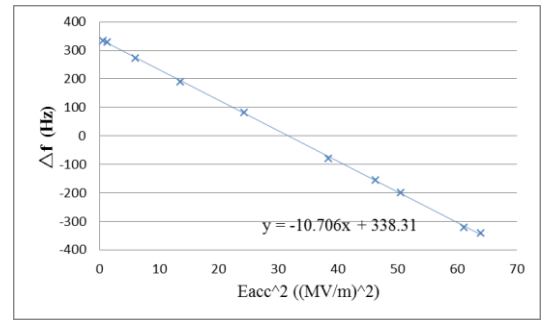

Fig.10. The Lorentz coefficient of 7# spoke012 cavity.

The Lorentz coefficient ($\sum_\mu k_\mu$ =-10) is so high that the frequency shift caused by Lorentz force is beyond several half bandwidths at 8MV/m.

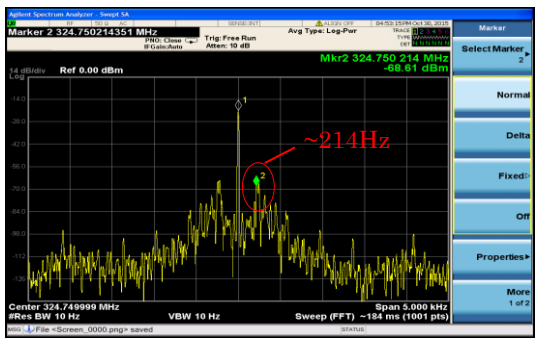

Fig.9. Frequency spectrum of pickup signal from the cavity when driven by a fixed-frequency source in the presence of ponderomotive effects.

In consideration of ponderomotive effects and microphonics, the frequency loop should provide more precise control to reject the electro-mechanical coupled resonance.

## 4.2 Feedback control of piezo tuner

When the amplitude and phase control loops are closed, we have compared the amplitude and phase stability of cavity field with feedback controller on and off. The results of the test are shown in Fig. 11 and Fig. 12. With PI feedback controller on, the amplitude stability of field has increased from ±0.9% to ±0.3% (peak to peak) and the phase stability of field has improved from ±3° to ±0.7°. Meanwhile, the max accelerating field rises from 5.5 MV/m up to 10 MV/m. The combination of fast feedback and slow feedback compensation successfully locked the cavity resonance to a fixed frequency but the long term drift is inevitable.

Another test has been done to verify the interactions between the two piezo actuator tuners. It demonstrated that they will not give rise to the multiple resonances at 6MV/m in 4 hours under proton beam test.

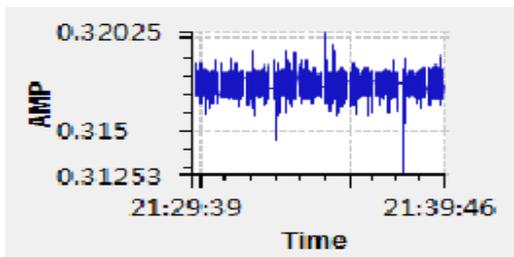 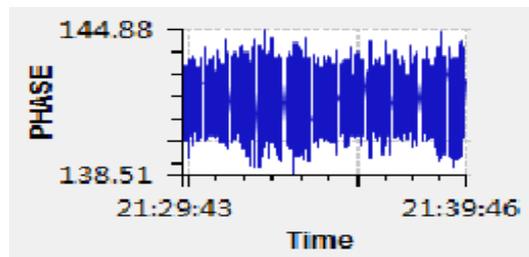

Fig.11. Amplitude (left) and phase (right) stability of field with feedback controller off.

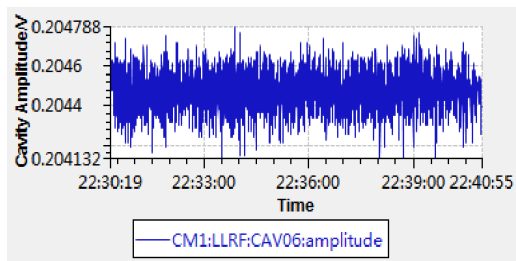 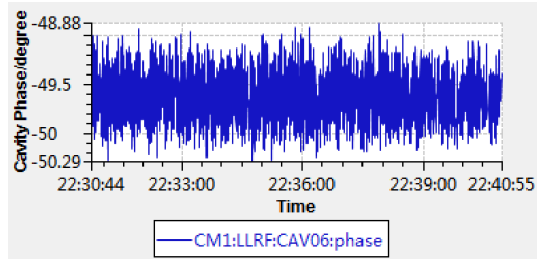

Fig.12. Amplitude (left) and phase (right) stability of field with feedback controller on.

## 5. Summary

This tuner control system has firstly been implemented to achieve the phase error of ±0.7°. It has been successfully applied to the CM1 of C-ADS injector I under proton beam 6MeV@1ms for pulse operation. In addition to this, the feedback control has been shown to be an effective strategy for reducing cavity detuning in a noisy vibrational environment. However, characteristics of cavity should be analyzed to understand the coupled electromechanical system and apply feedforward algorithms in order to design an optimal resonance controller.

*Many thanks to Prof. Y. Sun and Prof. G.W. Wang for their constant help and instruction during these years.*